\documentclass[aps,prl,reprint,floatfix]{revtex4-2}

\usepackage{amssymb}
\usepackage{amsmath}
\usepackage{amsfonts}
\usepackage{graphicx}

\newcommand{\ssec}[1]{\textbf{\emph{#1}}}

\newcommand{\FE}{\kappa}

\renewcommand{\vec}[1]{\boldsymbol{#1}}

\newcommand{\beq}{\begin{eqnarray}}
\newcommand{\eeq}{\end{eqnarray}}

\newcommand{\tr}{\text{Tr}}

\newcommand{\half}{\frac{1}{2}}

\newcommand{\rcite}[1]{Ref.~\onlinecite{#1}}

\newcommand{\rcites}[1]{Refs~\onlinecite{#1}}

\newcommand{\DFA}{\text{DFA}}

\newcommand{\GX}[1]{\text{GX#1}}

\newcommand{\Hx}{\text{Hx}}

\newcommand{\xrm}{\text{x}}
\newcommand{\crm}{\text{c}}
\newcommand{\Hrm}{\text{H}}
\newcommand{\xc}{\text{xc}}
\newcommand{\Hxc}{\text{Hxc}}
\newcommand{\Hc}{\text{Hc}}

\renewcommand{\sd}{{\text{sd}}}
\newcommand{\dd}{{\text{dd}}}

\newcommand{\ddA}{\text{ddA}}

\newcommand{\resp}{\text{resp}}
\newcommand{\coul}{\text{Coul}}
\newcommand{\WrespX}{W^{\resp}}
\newcommand{\Wresp}[1]{W^{#1,\resp}}
\newcommand{\Wcoul}[1]{W^{#1,\coul}}

\newcommand{\ExcGXSD}{E_{\xc}^{\text{LC}}}
\newcommand{\ExcGXDD}{E_{\Hc}^{\ddA}}

\newcommand{\EHx}{{\cal E}_{\Hx}}
\newcommand{\EH}{{\cal E}_{\Hrm}}
\newcommand{\Ex}{{\cal E}_{\xrm}}
\newcommand{\Ec}{{\cal E}_{\crm}}

\newcommand{\E}{{\cal E}}

\newcommand{\W}{{\cal W}}

\newcommand{\HInt}{U}

\newcommand{\pr}{^{\prime}}

\renewcommand{\vr}{\vec{r}}

\newcommand{\vrp}{\vec{r}\pr}

\newcommand{\wv}{\vec{w}}

\newcommand{\iket}[1]{|#1\rangle}

\newcommand{\ibraketop}[3]{\langle#1|#2|#3\rangle}
\newcommand{\ibkouter}[1]{|#1\rangle\langle#1|}

\newcommand{\iout}{\ibkouter}

\newcommand{\HF}{{\text{HF}}}

\newcommand{\up}{\mathord{\uparrow}}
\newcommand{\down}{\mathord{\downarrow}}

\newcommand{\Sgs}{\mathord{{}^1S_0}}
\newcommand{\Sts}{\mathord{{}^3S_1}}
\newcommand{\Ssx}{\mathord{{}^1S_1}}
\newcommand{\Sdx}{\mathord{{}^1S_2}}

\newcommand{\vh}{\hat{v}}
\newcommand{\nh}{\hat{n}}
\newcommand{\Th}{\hat{T}}

\newcommand{\Wh}{\hat{W}}

\newcommand{\Hh}{\hat{H}}

\newcommand{\SCE}{\text{SCE}}
\newcommand{\ESCE}{E_{\Hxc}^{\SCE}}

\newcommand{\Gammah}{\hat{\Gamma}}

\usepackage[normalem]{ulem}
\usepackage{color}
\usepackage{xcolor}

\definecolor{Mygrey}{gray}{0.80}
\definecolor{lteal}{rgb}{0.10,0.60,0.70}
\definecolor{dkred}{rgb}{0.40,0.10,0.00}
\definecolor{Red}{rgb}{0.80,0.00,0.00}
\definecolor{Navy}{rgb}{0.00,0.00,0.50}
\definecolor{Magenta}{rgb}{0.74,0.20,0.60}
\definecolor{Green}{rgb}{0.24,0.71,0.29}
\definecolor{Teal}{rgb}{0.00,0.50,0.50}
\definecolor{Blue}{rgb}{0.00,0.00,0.60}

\newcommand{\SM}[1]{\textcolor{dkred}{SMSec~#1}}
\newcommand{\SMC}[1]{\textcolor{dkred}{#1}}

\newif\iflite
\litetrue 

\iflite%
\newcommand\TG[2]{#2}

\else%
\newcommand\TG[2]{\textcolor{dkred}{#2}}

\fi

\newcommand\SwitchA[1]{}

\newcommand{\comment}[1]{}

\begin{document}
\title{\TG{}{State-specific density functionals for excited states from ensembles}}

\author{Tim Gould}\email{t.gould@griffith.edu.au}
\affiliation{Qld Micro- and Nanotechnology Centre, %
  Griffith University, Nathan, Qld 4111, Australia}
\author{Stephen G. Dale}
\affiliation{Qld Micro- and Nanotechnology Centre, %
  Griffith University, Nathan, Qld 4111, Australia}
\affiliation{Institute of Functional Intelligent Materials,
National University of Singapore, %
  4 Science Drive 2, Singapore 117544}
\author{Leeor Kronik}
\affiliation{Department of Molecular Chemistry and Materials Science, Weizmann Institute of Science, Rehovoth 7610000, Israel}
\author{Stefano Pittalis}
\affiliation{CNR-Istituto Nanoscienze, Via
  Campi 213A, I-41125 Modena, Italy}

\begin{abstract}
\TG{}{
We present a {\em first principles} strategy for developing state-specific density functional approximations for excited states.
We first clarify why approaches based on conventional ground state approximations \emph{miss} density-driven correlations, by considering excited state physics through the lens of ensemble density functional theory.
To solve this issue we gain insights on density driven correlations by exploiting the recently understood low-density limit of electrons in excited states.
The theory developments are then combined to produce a proof-of-concept excited state approximation that resolves urgent paradigmatic failures (double excitations, charge transfer excitations, piecewise linearity) of existing state-of-art density-functional approaches, directly from differences of self-consistent field calculations; i.e., $\Delta$SCF.
In light of its observed impressive performance, we conclude that the approach represents a major step toward {\em unified} and accurate modelling of neutral and charged excitations.}
\end{abstract}
\maketitle

\comment{
\TG{}{\bf Key (IMO) takeaways from the workshop, reviewer and chats with Stephen:
\begin{itemize}
    \item We have repeatedly failed to stress (including to experts at the workshop) that we do $\Delta$SCF DFT using EDFT as a framework to develop useful approximations;
    \item Our abstract over-emphasised GX24;
    \item We did not do a good job of explaining our EDFA development -- additions like ``response ansatz'' are there to address that;
    \item Especially, we over-explained `what' and under-explained `why';
    \item Our ddc derivation was overly complicated -- I took the derivation from the Perspective as a better template.
\end{itemize}
Hopefully all are better here but we ned to keep these points in mind during editing.}
}

\ssec{Introduction}:
The simulation and understanding of electronic structure has been transformed by inexpensive and useful density functional approximations (DFAs) ~\cite{Perdew2001-Jacob,Perdew2003,Kummel2008,Becke2014-Perspective} built on exact density functional theory (DFT) expressions and constraints.
Going beyond ground states usually requires time-dependent DFAs (TDDFAs) which, like their ground state counterparts, have transformed our understanding of excited states \cite{Ullrich2016}.
However, standard TDDFAs struggle to describe physically and technologically important processes, notably charge transfer~\cite{Maitra2017-CT} and double excitations.~\cite{Elliott2011}
Limitations may be overcome via specialised TDDFA-based approaches ~\cite{deWijn2008,Stein2009,Casanova2020}, but a generally applicable approach is as yet unavailable.

A large class of excitations involve stationary states and do not call for a truly time-dependent approach,~\cite{Matsika2018-ChemRev} which is strictly needed only for non-equilibrium physics. 
Hence, state-specific approaches which reuse regular DFT approximations (DFAs) are flourishing~\cite{Levi2020,Ivanov2021,Hait2021}.
However, predictions can be extremely sensitive to functional choice~\cite{Hait2021} because of non-systematic errors.
\TG{}{These issues arise, partly, from the fact that the central approximations were originally designed for ground states, so miss any specific electronic structure of excited states.}

\TG{}{An alternative path that is attracting growing attention is to develop approaches from ensemble DFAs (EDFAs) that are designed to directly target a specific state of interest (i.e. are state-specific) by incorporating first principles understanding of excited states~\cite{Filatov2015-Review,Cernatic2021,Gould2021-DoubleX,Gould2024-Stationary,Fromager2024-Stationary}.}
Modern EDFAs systematically and successfully address `typical' excited state problems~\cite{Filatov1999-REKS,Filatov2015-Double,Yang2017-EDFT,Gould2020-Molecules,Gould2021-DoubleX,Gould2022-HL}
and even multi-configurational problems like bond dissociation.~\cite{Filatov2015-Review}
The key advantage of EDFAs is that, \TG{}{because they incorporate excited state physics,} they \emph{upgrade} existing ground state DFAs for excited state predictions.

A significant difficulty that has been recognized only recently is that  EDFT introduces an additional form of ``density-driven'' (dd) correlations~\cite{Gould2019-DD,Fromager2020-DD,Gould2020-FDT}.
\TG{}{These terms do not appear in ground states, so are not captured by existing DFAs, yet must be accounted for in order to make accurate and general predictions.}

Originally, dd-correlations were brought forth by the realization that the individual auxiliary pure states forming the Kohn-Sham ensembles of EDFT do not necessarily need to reproduce the individual interacting densities~\cite{Gould2019-DD}. Rather,  what counts at the functional level is the ensemble-averaged density.
The definition of dd-correlations was revisited and generalized to handle more direct calculations and confirm its importance~\cite{Fromager2020-DD}.
Eventually~\cite{Gould2020-FDT}, a formally exact expression was derived to inspire approximations involving, explicitly, pure-state densities and density-like overlaps of different states~\cite{Gould2019-DD}, an aspect elaborated below.

This Letter derives, from first principles, a {\em so-far-elusive} practical approximation for dd-correlations\TG{}{, by first reframing existing understanding as an ansatz and then using the low-density limit of electrons~\cite{Gould2023-ESCE} to reintroduce physics missed by the ansatz.}
We thus produce a unified EDFA for ground (G) and excited (X) states, which we call GX24, that simultaneously and reliably addresses double and charge transfer excitations, as well as singlet-triplet splitting -- all neutral excitations of great relevance in photochemistry for which conventional TDDFAs struggle.
We \TG{further}{finally} demonstrate that GX24 is equally useful for charged excitations\TG{}{, before concluding}.

\TG{}{\ssec{Using EDFT to deduce excited state energies}:}
To begin, let us briefly summarise how EDFT lets us address excited states.
Seminal work by Gross, Oliveira and Kohn~\cite{GOK-1} (GOK) showed that ensembles of ground and excited states are subject to an extended variational (stationary) formalism. 
Specifically,
$\E^{\wv}
:=\min_{\{\Psi\}}\tr[\Hh\Gammah_{\Psi}^{\wv}]$,
where $\Gammah_{\Psi}^{\wv}=\sum_{\FE}w_{\FE}\iout{\Psi_{\FE}}$ is formed on orthonormal states, $\{\iket{\Psi_{\FE}}\}$, averaged by  weights $w_{\FE}$.
They also showed that $\E^{\wv}$ can be obtained via an ensemble density functional, $\E^{\wv}[n^{\wv}]$, of the ensemble density $n=\tr[\nh\Gammah^{\wv}]=\sum_{\FE}w_{\FE}n_{\FE}$~\cite{GOK-2}.
A recent reformulation of GOK theory, obtained by invoking the same weights for the interacting and the auxiliary symmetry-adapted KS states (denoted by the subscript `$s$'), yields~\cite{Gould2017-Limits,Gould2020-FDT}
\begin{align}
\E^{\wv}=&  \sum_\FE w_\FE \big[ T_{s,\FE} + \int n_{s,\FE}(\vr) v_{\rm ext}(\vr) d\vr + E_{{\rm Hxc},\FE} \big]\;,
\label{eqn:EE0DFT}
\end{align}
where $T_{s,\FE} = \ibraketop{\FE_s}{\Th}{\FE_s}$ is the state-specific KS kinetic energy, $n_{s,\FE} = \ibraketop{\FE_s}{\nh}{\FE_s}$ is its density, $v_{\rm ext}$ is the external potential (here, the scalar potential of the electric field due to the nuclei of a molecule), and $E_{{\rm Hxc},\FE} = E_{{\rm Hx},\FE} + E_{{\rm c},\FE}$, where $E_{{\rm Hx},\FE} = \ibraketop{\FE_s}{\Wh}{\FE_s}$ is the state-specific KS Hartree-exchange energy and $E_{{\rm c},\FE}$ is the state-specific correlation energy (elaborated below)~%
\footnote{Notice also that the KS  states, $\iket{\FE_s}$, and their properties are  implicitly weight-dependent and formally should be written as  $\iket{\FE_s^{\wv}}$. For notational simplicity we omit this weight dependence unless it is contextually essential.}.
\TG{}{Here, $\Th$ and $\Wh$ are the kinetic and interaction energy operators for electrons, respectively.
KS states in EDFT are necessarily~\cite{Gould2017-Limits,Gould2023-ESCE} non-interacting symmetry-adapted configuration-state functions (CSFs) which are a specific (yet \emph{finite}) combination of Slater determinants.}
 
\TG{}{CSFs do not always lead} to the usual determinant-based Hartree and Fock exchange separation~\cite{Pribram-Jones2014,Gould2017-Limits} \TG{}{that is central to the design and development of DFAs.
To resolve this issue, we follow \rcite{Gould2020-FDT} to instead define $E_{\xrm,\FE}:= \WrespX[\chi_{s,\FE}]$ as a functional of the KS (retarded density-density) response function, $\chi_{s,\FE}$, by invoking the fluctuation-dissipation theorem (FDT) through $\WrespX[\chi]:=\int \tfrac{-1}{\pi}\Im\int_{0}^{\infty}\chi(\vr,\vrp;\omega) d\omega - n[\chi](\vr)\delta(\vr-\vrp) \tfrac{d\vr d\vrp}{2|\vr-\vrp|}$.
This definition gives the usual Fock exchange for ground states, i.e. $\WrespX[\chi_s]=E_{\xrm}^{\text{Fock}}$, but can differ for excited states.
We will elaborate on the benefits of using the FDT for excited states near Eq.~\eqref{eqn:ExcCombo}.}

The ensemble Hartree energy, $\EH^{\wv} := \EHx^{\wv}-\Ex^{\wv} = \sum_{\FE} w_{\FE} E_{\Hrm,\FE}$, then captures any remaining~\cite{Gould2020-FDT} Coulombic Hx terms via
\begin{align}
    E_{\Hrm,\FE}  =  U[n_{s,\FE}] + 2\sum_{\FE'<\FE} U[n_{s,\FE\FE'}]\;.
    \label{eqn:EEH} 
\end{align}
Here, $U[q]=U[q^*]=\int \frac{d\vr d\vrp}{2|\vr-\vrp|} q(\vr)q^*(\vrp)$
is a Coulomb-like integral adapted to admit complex inputs like transition densities, $n_{s,\FE\FE'}=\ibraketop{\FE_s}{\nh}{\FE^{\prime}_s}$.
The sum in the above equation is over states $\iket{\FE'}$ that are lower in energy than $\iket{\FE}$. 
Degenerate orbitals sometimes require a more sophisticated treatment~\cite{Gould2021-DoubleX}. Here we focus on non-degenerate singlets to avoid complications.

\TG{}{Correlations (i.e. energy contributions missed by the KS state) can also be understood via the FDT~\cite{Gould2020-FDT}, by combining it with the adiabatic coupling-constant integration~\cite{Nagy1995}.}
The resulting ensemble correlation can be expressed as
$\Ec^{\wv}[n]=\sum_{\FE}w_{\FE} E_{\crm,\FE}=\sum_{\FE}w_{\FE}\int_0^1 (W_{\FE}^{\lambda}-W_{\FE}^0) d\lambda$.
Here, $W_{\FE}^{\lambda}:=\ibraketop{\FE^{\lambda}}{\Wh}{\FE^{\lambda}}$, where  $\iket{\FE^{\lambda}}$ is an eigen-state of $\Th+\vh^{\lambda}+\lambda\Wh$ with $v^0=v_s$, $v^1=v$ and $v^{\lambda}$ otherwise chosen to preserve the (ensemble) density.
The correlation energy is further resolved into: i) state-driven (sd) correlations,
\TG{}{$E_{\crm,\FE}^{\sd}:=\int_0^{1} (\Wresp{\lambda}_{\FE}-\Wresp{0}_{\FE}) d \lambda $, obtained by applying the FDT, $\Wresp{\lambda}_{\FE}\equiv \WrespX[\chi_{\FE}^{\lambda}]$, to interacting ($\chi_{\FE}^{\lambda}$) and non-interacting ($\chi_{\FE}^0=\chi_{s,\FE}$) response functions}; and ii) density-driven (dd) correlations,
\begin{align}
    E^{\dd}_{\crm,\FE}
    := &  \int_{0}^1 d\lambda (U[n_{\FE}^{\lambda}] - U[n_{s,\FE}])
    \nonumber\\&
  + 2 \sum_{\FE'<\FE} \int_{0}^1 d\lambda 
   (U[n_{\FE\FE'}^{\lambda}] - U[n_{s,\FE\FE'}])\;,
  \label{eqn:EEcDD}
\end{align}
involving Coulomb-like integrals with $n_{\FE\FE'}^{\lambda} =\ibraketop{\FE^{\lambda}}{\nh}{\FE^{\prime \lambda}}$ and $n_{s,\FE\FE'}\equiv n_{\FE\FE'}^{0}$.
\TG{}{We may also write this as $E_{\crm,\FE}^{\dd}:=\int_0^1 \Wcoul{\lambda}_{\FE} - \Wcoul{0}_{\FE} d\lambda$, in analogy to $E_{\crm,\FE}^{\sd}$.
The sd correlation (sdc) energy thus captures contributions \emph{within} the specific state, while the dd correlation (ddc) energy captures contributions \emph{between} non-interacting and interacting (transition) densities.}

\TG{}{\ssec{Modeling excited state energies via density functionals}:
Standard DFAs are devised for the lowest Slater determinant solution of each given spin-symmetry -- henceforth referred to as spin ground states (SGS). 
Our goal is to devise useful EDFAs for excited states.
Consistent with the seminal work of Kohn and Sham~\cite{KohnSham}, we treat $T_{s,\FE}$ and $E_{\Hrm,\FE}$ exactly via their excited state generalizations.
We then only need to model the unknown exchange-correlation (xc) energy term, $E_{\xc,\FE}=E_{\xrm,\FE}+E_{\crm,\FE}$, ideally by reusing existing SGS DFAs, $E_{\xc}^{\DFA}$, as much as possible.}

\TG{}{With modelling in mind we recognise that both $E_{\xrm,\FE}$ and $E_{\crm,\FE}^{\sd}$ involve response functions and are thus naturally paired.
Furthermore, the relation $E_{\xc}=\int_{0}^1 \Wresp{\lambda} d\lambda$ is exact in any SGS~\cite{Harris1974}.
We therefore  make an ansatz that all response physics can and \emph{should} be captured by standard DFAs, which leads to the approximation (see \rcite{Gould2021-DoubleX} for details) that exchange (x) and sdc approximations obey combination rules,}
\begin{align}
 E_{{\rm x},\FE}[n_\FE] + E^{\sd}_{{\rm c},\FE}[n_\FE] \approx \sum_P C_P^{\FE} E^{\DFA}_{\rm xc}[n_P]\;.
 \label{eqn:ExcCombo}
\end{align}
\TG{}{
Here, $n_P$ are  densities of SGS Slater determinants, $\iket{\Phi_P}$, while $C_P^{\FE}$ are coefficients such that 
$\sum_P C^{\FE}_P=1$, $\sum_P C^{\FE}_P n_P=n_{\FE}$ and $\sum_P C^{\FE}_P E_{\xrm,P}=E_{\xrm,\FE}$%
~\footnote{\TG{}{The combination rule relationship follow from the expressions derived in {\rcite{Gould2021-DoubleX}} that the exchange energy depends linearly on orbital occupation factors, as do densities, for fixed orbitals and orbital energies.
Eq.~\eqref{eqn:ExcCombo} yields reasonable excited state energies~\cite{Gould2021-DoubleX,Gould2022-HL} despite only partially capturing differences between same-spin and opposite-spin correlations.Thus, for any target density expressible via a combination law of densities, it follows that the exchange energy obeys the same combination law.
It is an approximation (i.e. combination law$\to$rule) for correlations because the interacting response function, $\chi^{\lambda}$, should \emph{also} reflect differences between same-spin and opposite-spin correlations, whereas the non-interacting response function, $\chi_s$, used to derive the combination rule does not.
The Hartree energy and density-driven correlation energy capture some, but not all, of these differences.}}.
The response ansatz and Eq.~\eqref{eqn:ExcCombo} allow well-developed SGS infrastructure to be repurposed for the \emph{response} physics of excited states.}
The outstanding challenge is thus with the density-driven correlations in Eq.~\eqref{eqn:EEcDD}.

\ssec{Modeling density-driven correlations (ddc)}:
The crux of the problem with ddc is that $E^{\dd}_{{\rm c},\FE}[n] $ \TG{}{is exactly zero in the SGS that are at the center of all popular DFA constructions.
All terms in Eq.~\eqref{eqn:EEcDD} disappear because the KS state has the correct density at all $\lambda$ (i.e. $n_s=n^{\lambda}=n$) and no relevant eigenstate can be found that is lower in energy than the SGS~\footnote{\TG{}{Some SGS are excited states.
However, any eigenstate lower in energy must, by definition, have a different spin-symmetry; and it follows that the transition density must be exactly zero.
Such states are therefore irrelevant.}}.
Consequently, existing models do not know about and thus \emph{cannot} describe ddc.}

For higher energy states we enter a largely unexplored territory.
Let us first consider the situation in which we are able to select a specific state such that $n_{\FE,s} = n_{\FE}^{\lambda} = n_{\FE}$ is constant along the adiabatic connection. 
Then, the first term in Eq.~\eqref{eqn:EEcDD} vanishes. But, in general, $n_{s,\FE\FE'} \neq n^{\lambda}_{\FE\FE'}$ leaving us with \TG{}{\emph{non-vanishing}},
\begin{align}
E^{\dd}_{{\rm c},\FE}[n]  = 2 \sum_{\FE'<\FE} \int_{0}^1
(U[n_{\FE\FE'}^{\lambda}]-U[n_{s,\FE\FE'}]) d\lambda   \;.
\label{eqn:EcDD_FDT}
\end{align}
\TG{}{A conventional DFA description, e.g. via Eq.~\eqref{eqn:ExcCombo} or a similar model, of the excited state will miss this physics.}

\TG{}{Evaluating Eq.~\eqref{eqn:EcDD_FDT} involves the hitherto impervious task of modeling the variation of transition densities, $n_{\FE\FE'}^{\lambda}$, for all $\lambda$ in the integration path. 
As a first step toward an approximation, we rewrite the conjoint Hartree, exchange and correlation (Hxc) energy,
\begin{align}
E_{\Hxc,\FE}=\int_{0}^1 W_{\FE}^{\lambda}d\lambda
:=\int_{0}^1 \big(\Wresp{\lambda}_{\FE} + \Wcoul{\lambda}_{\FE} \big) d\lambda\;,
\label{eqn:EcW}
\end{align}
in terms of response ($\Wresp{\lambda}_{\FE}$) and Coulombic ($\Wcoul{\lambda}_{\FE}$) contributions.
Our ansatz associates the response contributions with x and sdc [Eq.~\eqref{eqn:ExcCombo}], i.e. $\int_{0}^1 \Wresp{\lambda}_{\FE}d\lambda\approx\sum_P C_P^{\FE}E_{\xc}^{\DFA}[n_P]$.
It follows that the Hartree [Eq.~\eqref{eqn:EEH}] and ddc [Eq.~\eqref{eqn:EcDD_FDT}] contributions are associated with $\int_{0}^1 \Wcoul{\lambda}_{\FE} d\lambda=E_{\Hrm,\FE}+E_{\crm}^{\dd}$.
}

\TG{}{
The function $W_{\FE}^{\lambda}$ is unknown, in general, but was recently shown~\cite{Gould2023-ESCE} to obey $W_{\FE}^{0}=E_{\Hx,\FE}$ (the high-density limit of electrons) and $W_{\FE}^{\infty}=\ESCE[n_{s,\FE}]$ (the low-density limit), where $\ESCE[n]$ is the energy functional for \emph{ground state} strictly-correlated elections~\cite{Seidl1999}.
Interpolating between the known limits leads to $W_{\FE}^{\lambda}[n]:=[1-f_{\FE}[n](\lambda)] E_{\Hx,\FE} + f_{\FE}[n](\lambda) \ESCE[n_{s,\FE}]$, where $f_{\FE}[n](\lambda)$ is an unknown monotone function obeying $f_{\FE}[n](0)=0$ and $f_{\FE}[n](\infty)=1$ for all $\FE$ and $n$.
Integrating over $\lambda$ therefore yields the approximation,
\begin{align}
E_{\Hxc,\FE}\approx (1-\xi)E_{\Hx,\FE} + \xi \ESCE[n_{s,\FE}]\;,
\label{eqn:TWL} 
\end{align}
after assuming that $f^{\lambda}\approx f_{\FE}[n](\lambda)$ and thus $\xi=\int_0^1 f^{\lambda} d\lambda$ are independent of $\FE$ and $n$.
While this assumption may seem drastic, it is reasonable for our goal of approximating ddc because it applies, in practice, at $O(\lambda^2)$.
An illustrative geometric derivation for singlet-triplet gaps, and more details about the model, are provided in \SMC{Supplementary Material Section I} (\SM{I}).}

\TG{}{
Next, we expand $E_{\Hxc,\FE}\approx (1-\xi)(E_{\Hrm,\FE} + E_{\xrm,\FE})+ \xi (U[n_{s,\FE}] + E_{\xc}^{\SCE}[n_{s,\FE}])$ in terms of its constituent parts.
Here, we divided $\ESCE[n]:=U[n]+E_{\xc}^{\SCE}[n]$ into Coulombic (just $U[n]$ because it is a ground state functional) and xc parts.
We then recognise that for any SGS, $\iket{\Phi_P}$, we obtain $E_{\Hrm,P}=U[n_{s,P}]$ and thus $E_{\xc,P} \approx (1-\xi)E_{\xrm,P} + \xi E_{\xc}^{\SCE}[n_{s,P}]$ captures only the explicit `x' and `xc' terms.
It follows that these terms are SGS physics and the response ansatz thererfore associates them with $\Wresp{\lambda}_{\FE}$ and Eq.~\eqref{eqn:ExcCombo}.}

\TG{}{
That leaves us with only the `H' and `$U$' terms to be associated with $\Wcoul{\lambda}_{\FE}$, so that $\int_0^1 \Wcoul{\lambda}_{\FE} d\lambda = (1-\xi)E_{\Hrm,\FE} + \xi U[n_{\FE}]$.
The dd approximation (ddA) is thus $E_{\crm,\FE}^{\dd}\approx E_{\crm,\FE}^{\ddA}:=\xi (U[n_{\FE}]-E_{\Hrm,\FE})$ in general and,
\begin{align}
E_{\crm,\FE}^{\ddA}:=  -2\xi \sum_{\FE'<\FE} U[n_{s,\FE\FE'}]
\label{eqn:EcDD}
\end{align}
for non-degenerate (and some degenerate, e.g. the two-fold degenerate case later reported in Table~\ref{tab:EDFA}) excitations%
~\footnote{
\TG{}{General degeneracies must by treated using energy \emph{levels}, not individual states, which can lead to contributions from the first term in Eq.~\eqref{eqn:EEcDD} and thus necessitates use of Eq.~\eqref{eqn:EEcDDw}.}}.
Repeating all steps for ensembles (i.e. beginning from $\W^{\wv,\lambda}=\tr[\Gammah^{\wv,\lambda}\Wh]$) yields,
\begin{align}
\Ec^{\wv,\ddA}:= \xi ( U[n^{\wv}] - \EH^{\wv}[n^{\wv}] )\;,
\label{eqn:EEcDDw}
\end{align}
where $U[n^{\wv}]$ follows from the SCE limit being independent of the weights, $\wv$, except via the density $n^{\wv}$~\cite{Gould2023-ESCE}.}


\newcommand{\alphaopt}{\tfrac{3}{8}}
\newcommand{\gammaopt}{0.2}
\newcommand{\xiopt}{0.32}
\newcommand{\STPre}{1.36} 

\begin{table}[t!]
\begin{ruledtabular}\begin{tabular}{lc}
State & 
$\sum_P C^{\FE}_P \ExcGXSD[\rho_P] \nonumber + \sum_{ia} D^{\FE}_{ia}\ExcGXDD[\phi_i,\phi_a]$
\\\hline
$\Sgs$ & $\ExcGXSD[\rho_{\Sgs}]$
\\
$\Sts$ & $\ExcGXSD[\rho_{\Sts}]$
\\
$\Ssx$ & $\ExcGXSD[\rho_{\Sts}] + \ExcGXDD[\phi_h,\phi_l]$
\\
$\Sdx$ & $2\ExcGXSD[\rho_{\Sts}] - \ExcGXSD[\rho_{\Sgs}] + \ExcGXDD[\phi_h,\phi_l]$
\\
$^1D_2$ &
$\ExcGXSD[\rho_{\Sts;l_1}] + \ExcGXSD[\rho_{\Sts;l_2}] - \ExcGXSD[\rho_{\Sgs}]$
\nonumber\\
& $+ \half(\ExcGXDD[\phi_h,\phi_{l_1}]+ \ExcGXDD[\phi_h,\phi_{l_2}])+ \ExcGXDD[\phi_{l_1},\phi_{l_2}]$
\end{tabular}\end{ruledtabular}
\caption{Ingredients for GX24.
\TG{}{$\ExcGXSD$ and $\ExcGXDD$} are defined in Eq.~\eqref{eqn:ExcHcGX24}, and $\rho_{\Sgs}$ and $\rho_{\Sts}$ are 1RDMs for singlet- and triplet- SGSs.
$^1S_2$ and $^1D_2$ are double excitations ($h^2\to l^2$ and $h^2\to l_1l_2$) to $l$ and degenerate $l_{1,2}$, respectively.
\label{tab:EDFA}}
\end{table}

\ssec{The GX24 approximation}:
\TG{}{We are now ready to put the model into practice, by producing a proof-of-concept excited state approximation that pairs the density-driven correlation model of Eq.~\eqref{eqn:EcDD} with an existing long-range corrected (LC) hybrid DFA model that is known to facilitate the description of charge transfer excitation~\cite{Iikura2001}.
The resulting ground (G) and excited state (X) ``GX24'' has different energy expressions for different excited states, which can usually be expressed as,
\begin{align}
E_{\Hxc,\FE}^{\GX24}:=&T_{s,\FE} + \int n_{\FE} v d\vr + U[n_{\FE}]
+ \sum_P C^{\FE}_P \ExcGXSD[\rho_P]
\nonumber\\& 
+ \sum_{ia} D^{\FE}_{ia}\ExcGXDD[\phi_i,\phi_a]\;,
\label{eqn:EGX24}
\end{align}
Here, the first sum is from combination rules and the second captures any missing Hartree terms and the ddc approximation, where $D_{ia}^{\FE}$ depend on the specific excitation of interest%
~\footnote{\TG{}{We note that the expression may sometimes require modification to deal with degenerate energy levels or other kinds of ensembles like the fractional case considered later.
For example, terms $\propto [aa|aa]$ or $\propto [aa|bb]$ may also appear.
All terms can be determined by first evaluating $E_{\Hx,\FE}$ and $E_{\xrm,\FE}$, then using them to determine $E_{\Hrm,\FE}$ and $E_{\crm,\FE}^{\dd}$}.}
-- some important examples are provided in Table~\ref{tab:EDFA}, e.g. the fourth row yields,
$E^{\GX24}_{\Sdx}= T_{s,\Sdx} + \int n_{\Sdx} v d\vr + U[n_{\Sdx}]
+ 2\ExcGXSD[\rho_{\Sts}] - \ExcGXSD[\rho_{\Sgs}] + \ExcGXDD[\phi_h,\phi_l]$
as the energy of a non-degenerate double-excited state, $\Sdx$.
Terms based on $\ExcGXSD$ involve spin-resolved one-body reduced density matrices (1RDMs), $\rho_P$, of SGS Slater determinants.
$\ExcGXDD=2(1-\xi)[hl|lh]$ invokes electron repulsion integrals, $[ij|kl]=\int\phi_i(\vr)\phi_j^*(\vr)\phi_k(\vrp)\phi_l^*(\vr)\tfrac{d\vr d\vrp}{|\vr-\vrp|}$.}

\TG{}{More specifically, GX24 invokes
\begin{subequations}\begin{align}
&\ExcGXSD[\rho]:=E_{\xrm}^{\HF}[\rho]
 + E_{\crm}^{\text{PBE}}[n_{\up},n_{\down}]
 \nonumber\\&~~~~~~~~
 + \tfrac58 \big(
E_{\xrm}^{\text{HJS}_{\gammaopt}}[n_{\up},n_{\down}]
- E_{\xrm}^{\HF_{\gammaopt}}[\rho] \big)
\;,
\label{eqn:GX24a}
\\
&\ExcGXDD[\phi_h,\phi_l]:=2(1-\xiopt)[hl|lh]=\STPre[hl|lh]\;.
\label{eqn:GX24b}
\end{align}\label{eqn:ExcHcGX24}\end{subequations}
Here, `PBE' is the Perdew-Burke-Ernzerhof generalized-gradient approximation ~\cite{DFA:pbepbe} and `HJS' is the Henderson-Janesko-Scuseria PBE-like form~\cite{Henderson2008} for short-range exchange.
Energy expression are then solved self-consistently via restricted orbital theory for \emph{each} state of interest; and excitation energies are obtained by taking differences of self-consistent energies (i.e. $\Delta$SCF).
Implementation details, including \rcites{Broadway_v100,psi4-1,psi4-2,pyscf}, are in \SM{II}.
The choice of short-range HF exchange fraction, $\alpha=\alphaopt$, and range-separation parameter, $\gamma=\gammaopt$~Bohr$^{-1}$, to range-partition of the Coulomb interaction are justified in \SM{III}.}

\TG{}{
For our proof-of-concept we use constant $\xi:=\xiopt$.
The value is determined by optimizing excitation energies over EX22~\footnote{EX22 is composed of 21 low-lying excitation processes from QuestDB~\cite{Quest1,Quest2,Quest3} plus the degenerate double excitation of diatomic BH~\cite{Gould2021-DoubleX}}, a benchmark set of 22 excitation energies that capture a diversity of physical and chemical behaviours.
Details and discussion of EX22 and GX24 (including ground state tests) are in \SM{III}.
}

\begin{figure}[t!]
\includegraphics[width=\linewidth]{{{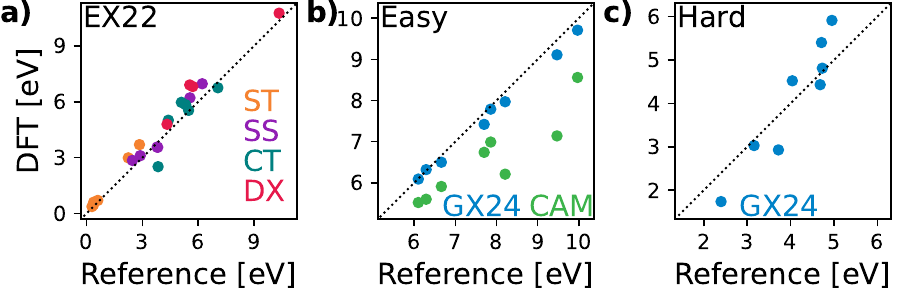}}}\\
\caption{
{\bf a)} Excitation energies ($y$ axis) as a function of reference values ($x$ axis) for EX22 \TG{}{using GX24.}
Colors indicate type of excitation.
{\bf b)} GX24 (blue) and CAM-B3LYP (green) energies versus reference energies for `easy' excitation energies.
{\bf c)} Like {\bf b)} but for `hard' double excitations that TDDFT cannot tackle.
Dotted lines indicate perfect agreement.
\label{fig:EX22}}
\end{figure}

\ssec{GX24 predictions}:
Figure~\ref{fig:EX22}a reveals that Eq.~\eqref{eqn:ExcHcGX24} performs extremely well for $\Sgs \to \Ssx$ (SS)
and $\Sts\to\Ssx$ (ST) transitions in EX22, with impressive mean absolue errors (MAE) of 0.19~eV and 0.04~eV, respectively.
It is worse, yet still respectable (in the absence of system-specific optimal tuning \cite{Stein2009}) for difficult charge-transfer (CT, MAE 0.58~eV) and does extremely well for the notorious case of double excitations (DX, 0.35~eV).
Using only the sd-EDFA [Eq.~\eqref{eqn:GX24a}] increases errors for all four categories, overall by 83\%, highlighting the importance of the dd-EDFA.
\TG{}{
Extended theory, technical details of all calculations, and tabulated data are in \SM{IV}.}

\TG{}{Further testing reveals strong performance outside \comment{ground states and} EX22.
Figure~\ref{fig:EX22}b shows results for `easy' lowest-lying singlet-singlet transitions of four small atoms, with a focus on Rydberg-like excitations for which very high quality theoretical reference data are available~\cite{Quest1}; Figure~\ref{fig:EX22}c extends the tests to `hard' problems with a substantial `double excitation' character~\cite{QuestDB-Dbl24}.
GX24 has a MAE of 0.18~eV across all eight `easy' excitations and a respectable MAE of 0.50~eV for the eight `hard' ones.
We compare with Tamm-Damcoff approximated CAM-B3LYP~\cite{DFA:cam-b3lyp} as representative of a well-utilized and generally effective~\cite{Sarkar2021,Hall2023} TDDFT method.
CAM-B3LYP is worse on easy excitations (Figure~\ref{fig:EX22}b) and very poor for the three accessible hard ones (which are underestimated by $1.5$--$3.4$~eV so do not fit in Figure~\ref{fig:EX22}c).
Other TDDFAs may out-perform GX24 on single excitations but, unlike GX24, they \emph{cannot} capture all of the double excitations.}

\begin{figure}[b!]
\includegraphics[width=\linewidth]{{{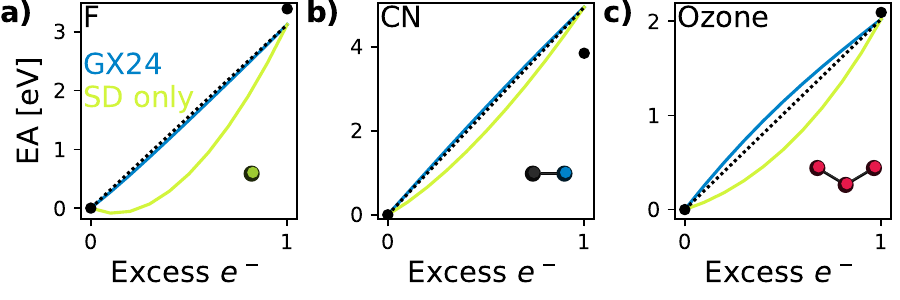}}}\\
\caption{
Electron affinity of F ({\bf a}), CN ({\bf b}) and ozone ({\bf c}) at fractional charges.
The lime line is GX24 without the dd-EDFA, while the blue line includes all terms.
The black dotted line is the theoretically exact linear behaviour and black dots are experimental values.~\cite{Blondel1989,Bradforth1993,Novick1979} 
\label{fig:Frac}}
\end{figure}

Our final test steps outside the realm of neutral excitations to address \emph{fractionally charged} excitation results.
These excitations involve ground states, but reproducing non-integer electron number requires an ensemble treatment~\cite{Perdew1982} and thus gives rise to dd correlations%
~\footnote{Note, in the ensemble case there are no transition densities to worry about. Rather, the $\lambda=0$ limit yields a weighted average of electrostatic energies, $(1-q)\HInt[n_N]+q\HInt[n_{N+1}]$ of $N$ and $N+1$ electron systems, whereas the $\lambda=\infty$ limit yields the electrostatic energy of the average, $\HInt[(1-q)n_N+qn_{N+1}]$, where $n_{N+1}=n_N+n_h$.
The dd correlation energy is a fraction of the difference.}.
Figure~\ref{fig:Frac} shows the electron affinity of F, CN, and ozone (O$_3$) at fractional charges, which in exact theory should be linear as a function of excess charge~\cite{Perdew1982}.
\TG{}{Applying the sd-EDFA part of GX24 (i.e. Eq.~\eqref{eqn:GX24a} adapted to known fractional charge combination rules~\cite{Gould2013-LEXX,Kraisler2013,Kraisler2014}) but omitting the dd-EDFA leads to substantial non-linearities (lime curves in the figure).
Eq.~\eqref{eqn:EEcDDw} yields $\Ec^{\ddA}(q)\approx -\xi\tfrac{q(1-q)}{2}[hh|hh]$ as the appropriate adaptation of Eq.~\eqref{eqn:GX24b}, where $\xi=\xiopt$, $q$ is the excess electron charge and $\phi_h$ is the frontier orbital.}
Including $\Ec^{\ddA}(q)$ (blue curves) largely restores linearity in all cases, to yield excellent qualitative and quantitative results for all charges.

\ssec{Conclusions}:
\TG{}{By systematic first principles modelling, and a transparent series of mathematical steps,} we derived a simple yet effective approximation, Eq.~\eqref{eqn:EcDD}, for density-driven correlations -- a central quantity for dealing with excited states via ensemble density functionals.
\TG{}{Eq.~\eqref{eqn:EcDD} is explicitly derived to complement a response ansatz, that lets us repurpose existing ground state DFA developments via Eq.~\eqref{eqn:ExcCombo}.}

\TG{}{We then used the model to produce a proof-of-concept EDFA, GX24 [Table~\ref{tab:EDFA} and Eqs~\eqref{eqn:EGX24}--\eqref{eqn:ExcHcGX24}], that successfully captures a variety of important ground- and excited state physics within a unified and adaptable $\Delta$SCF framework.}
Despite its simple form, the performance of GX24 stands beyond the reach of state-of-art time-dependent DFAs for addressing \TG{}{difficult excitations of relevance to photochemistry, including double and charged excitations}.
GX24's performance on charged excitations is particularly astonishing, given that it was derived and normed on neutral excitations.

\TG{}{Our model provides an effective framework for further improvements because simplifying assumptions in the present work can be relaxed.
Notably:
i) the modeling of the state-driven correlation physics can be refined to accommodate correlation differences between same- and oppositie-spin electrons and/or explicit excited state contributions~\cite{Gould2024-cofe};
ii) the LC-DFA used in Eq.~\eqref{eqn:GX24a} can be improved in form or via empiricism;
iii) parameters, including $\xi$, are system and state-dependent, which can be accommodated in approximations;
iv) Eq.~\eqref{eqn:EGX24} can be applied to other types of excitations, including challenging degenerate energy levels.}

\TG{}{Proof-of-concept GX24 already performs comparably to popular approximations on single excitations, yet can also address double excitations and charged excitations.
Given the demonstrated success, and the broad scope for further improvements, we believe that our work has significant potential to transform the way low-lying excited state properties are computed and understood.
Especially, accurate prediction of double excitations at DFT cost paves the way to understanding their role in photochemistry, biophysics and opto-electronics.}

\vspace{5mm}

\acknowledgments
TG, SGD, and LK were supported by an Australian Research Council (ARC)
Discovery Project (DP200100033).
TG was supported by an ARC Future Fellowship (FT210100663). LK was supported by the Aryeh and Mintzi Katzman Professorial Chair and the Helen and Martin Kimmel Award for Innovative Investigation.
SGD was supported by the Ministry of Education, Singapore, under its Research Centre of Excellence award to the Institute for Functional Intelligent Materials. Project No. EDUNC-33-18-279-V12.
Computing resources were provided by the Australian National Computing Merit Application Scheme (NCMAS sp13).

\end{document}